\documentclass{ptapap}
\usepackage{amsmath}
\usepackage{amssymb}
\usepackage{subcaption}

\usepackage[colorlinks=true,linkcolor=blue,citecolor=blue,urlcolor=blue]{hyperref}

\author{Anjitha John William}[cft] 
\author{Priyanka Jalan}[cft]
\author{Maciej Bilicki}[cft]
\author{Wojciech A. Hellwing}[cft]
\affil[cft]{Center For Theoretical Physics, Aleja Lotników 32/46, 02-668 Warsaw, Poland}

\title{Deep learning based photometric redshifts for the Kilo-Degree Survey bright galaxy sample}
\headtitle{Deep-learning photometric redshifts for KiDS DR4 bright galaxies}

\begin{document}

\maketitle

\begin{abstract}

In cosmological analyses, precise redshift determination remains pivotal for understanding cosmic evolution. However, with only a fraction of galaxies having spectroscopic redshifts (spec-$z$s), the challenge lies in estimating redshifts for a larger number. To address this, photometry-based redshift (photo-$z$) estimation, employing machine learning algorithms, is a viable solution. Identifying the limitations of previous methods, this study focuses on implementing deep learning (DL) techniques within the Kilo-Degree Survey (KiDS) Bright Sample for more accurate photo-$z$ estimations. Comparing our new DL-based model against prior `shallow' neural networks, we showcase  improvements in redshift accuracy. Our model gives mean photo-$z$ bias $\langle \Delta z\rangle= 10^{-3}$ and scatter $\mathrm{SMAD}(\Delta z)=0.016$, where $\Delta z = (z_\mathrm{phot}-z_\mathrm{spec})/(1+z_\mathrm{spec})$. This research highlights the promising role of DL in revolutionizing photo-$z$ estimation.
\end{abstract}

\section{Introduction}

Redshift is a basic quantity in cosmological analyses, serving as one of the indicators of galaxy distances. Its precise determination enables the mapping of the three-dimensional large-scale structure of the Universe and facilitates our comprehension of cosmic evolution. Achieving sub-percent accuracy in redshift measurements is exclusively feasible through spectroscopy. In obtaining spectroscopic redshifts (spec-zs), we initially capture the spectrum of a celestial source and subsequently identify the shift in spectral lines relative to their rest frame. 

Due to expensive telescope time needed, spec-$z$s are only attainable for a fraction of galaxies. However, a viable alternative is the estimation of redshifts through photometric measurements.  Such photometric redshifts (photo-$z$s) may not provide the same level of redshift accuracy and precision as spec-$z$s; however, they have proven indispensable in the era of massive imaging surveys. Such redshift estimates rely on the correlation between observed galaxy photometry and the true redshift \citep[e.g.][]{Baum1957}. This proceeding focuses on estimating this correlation using machine learning (ML) algorithms.

Machine learning-based techniques for photo-z estimation include supervised learning \citep{Wadadekar2005, Collister2004, Oyaizu2008}, unsupervised learning \citep{Way2012}, k-nearest neighbors \citep{Graham2018}, ensemble learning and Gaussian processes \citep{Way2006, Bonfield2010}, mixed density networks \citep{Ansari2021} and finally, deep neural networks \citep{Hoyle2015, Disanto2018}. In supervised ML techniques, the algorithm establishes an empirical relationship between observed quantities and corresponding labels through training on appropriately labeled data. The main challenge and limitation of these methods lie in extrapolating results beyond the representative training set.

Deep learning (DL) based photo-$z$ derivation is a promising technique among other supervised ML techniques \citep[e.g.][]{Menou2019, Dey2022,Li2022, Treyer2023}. 
In such frameworks, the term `deep' signifies that the models typically employ intricate, multi-layer architectures. Deep learning methods, specifically convolutional neural networks (CNNs), are selected for predicting photo-zs due to their exceptional ability to discern patterns in data, particularly images. 

In this work, we present a deep learning model for photometric redshift estimation of bright galaxies in the Kilo-Degree Survey \citep[KiDS,][]{deJong2013}, a multiband imaging survey covering about 1350 deg$^2$ of the sky. Following up on earlier KiDS work by \cite{Li2022}, we applied the DL models for photo-z estimation within the flux-limited `KiDS Bright Sample', encompassing all galaxies from KiDS with 
$r < 20$ mag \citep{Bilicki2021}. We compare our derivations with the previously released KiDS bright sample photo-zs obtained with `shallow' neural networks 
from the public package  ANNz2 \citep{Sadeh2016}. Our new model improves over those previous results and gives great perspectives for the final KiDS data release.

\section{Input Data}

This section details the datasets used in this study. The training sample required for the neural network is KiDS-DR4 four-band (u, g, r, and i) optical galaxy images and their 9-band magnitudes, with their corresponding labels of true (spectroscopic) redshifts from the Galaxy And Mass Assembly \citep[GAMA,][]{Driver2011} survey.

\subsection{Images and Photometry}
KiDS is an extensive optical survey conducted by the European Southern Observatory (ESO) at the VLT Survey Telescope \citep[VST,][]{Capaccioli2011}, which utilizes the OmegaCAM CCD mosaic camera, boasting an impressive 268 million-pixel focal plane. It is positioned at the ESO Paranal Observatory in Chile. The VST is an alt-azimuth mounted telescope based on a modified Ritchey-Chrétien design. KiDS captured images in four distinct broad bands ($ugri$) and encompasses an area of 1350 square degrees within the extragalactic sky.

We have used KiDS Data Release 4 \citep[DR4,][]{Kuijken2019}\footnote{\url{https://kids.strw.leidenuniv.nl/DR4}} co-added images that had been calibrated for both astrometry and photometry, ensuring a consistent pixel scale of 0.2 arcseconds. We have also incorporated the nine-band magnitudes of galaxies. 
These latter join original KiDS $ugri$ measurements 
with  
those from the VIKING \citep{Edge2013} survey, encompassing five near-IR bands ($ZYJHK_s$). 

\subsection{GAMA Spectroscopic Data}
The true (spectroscopic) redshifts used for training and testing 
are obtained from the 
GAMA dataset. This survey, conducted across five fields, of which four are fully within the KiDS footprint,  
covers a total area of approx.~286 square degrees. The data were obtained using the AAOmega fiber-fed spectrograph facility mounted on the 3.9-meter Anglo-Australian Telescope \citep{Driver2011}. GAMA gives spectra, redshifts, their quality, and other ancillary information. 
KiDS-DR4 fully overlaps with the G09, G12, G15 and G23 fields. In this work, we have used the GAMA-II spectroscopic redshift catalog of galaxies from the GAMA DR4  \citep{Driver2022}\footnote{\url{http://www.gama-survey.org/dr4}}.

We have identified galaxies within KiDS tiles that overlap between the GAMA and KiDS-bright sample based on their sky coordinates. Subsequently, we created cutouts for these galaxies, positioning each at the center of the cutout with dimensions of $7.2''\times7.2''$. The final cutout catalog comprises $\sim136$k galaxies in the equatorial field, with $\sim95$k utilized for model training. Model testing was conducted on $\sim20$k galaxies present in both the equatorial field and KiDS bright sample.

\subsection{KiDS-DR4 Bright Galaxy Sample}
The KiDS-bright galaxy sample \citep{Bilicki2021} is a collection of KiDS galaxies selected based on their flux and free from artifacts etc. 
Specifically, it includes galaxies with a magnitude limit of $r_\mathrm{auto} < 20$ mag, where ‘auto’ stands for SExtractor-derived estimate of the total flux via automatic aperture photometry. 

\section{Methodology}
\label{methodology}

Convolutional Neural Networks (CNNs), a subtype of Artificial Neural Networks (ANNs), excel in addressing computer vision challenges like image detection and classification. The operational principles of CNNs draw inspiration from the human neural system, mirroring the functionality of neurons \citep{McCulloch1943}. Similar to biological neurons that receive input and transmit electrochemical signals, CNNs employ artificial neurons to process information. The importance of CNNs becomes apparent through their widespread application in image recognition tasks, as highlighted by e.g. \cite{LeCun1998}. The adaptability and efficacy of CNNs make them a key technology in the field of computer vision, contributing significantly to advancements in image-based tasks. Our model includes a special CNN architecture called Inception \citep{Szegedy2015}. Our approach employs CNNs to forecast the photo-zs for the KiDS-bright sample, treating it as a regression problem. To label the training set, we used the spectroscopic GAMA data detailed earlier.

\begin{figure*}[!hb]
  \centering
  \begin{minipage}{0.5\textwidth}
    \centering
        \includegraphics[width=0.9\textwidth]{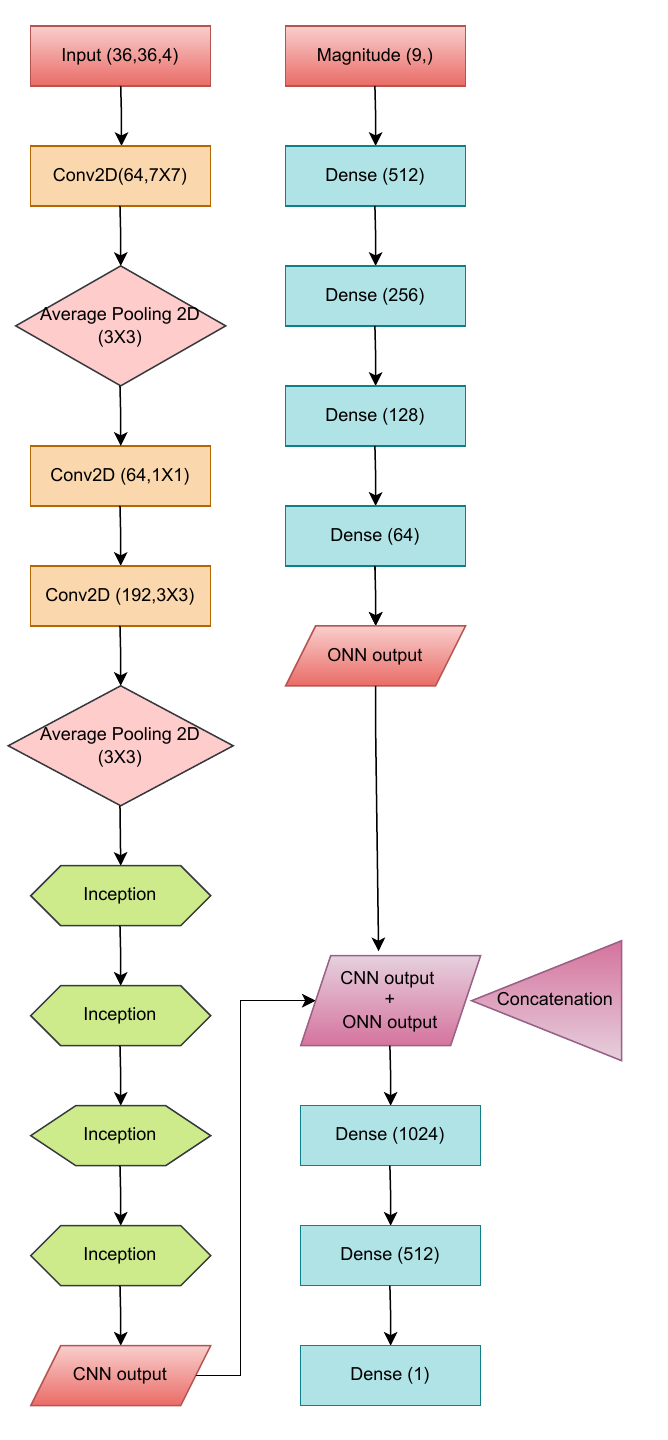}
    \subcaption{Architecture of the Combined model.\label{fig:model}}
  \end{minipage}%
  \begin{minipage}{0.5\textwidth}
    \centering
        \includegraphics[width=0.85\linewidth]{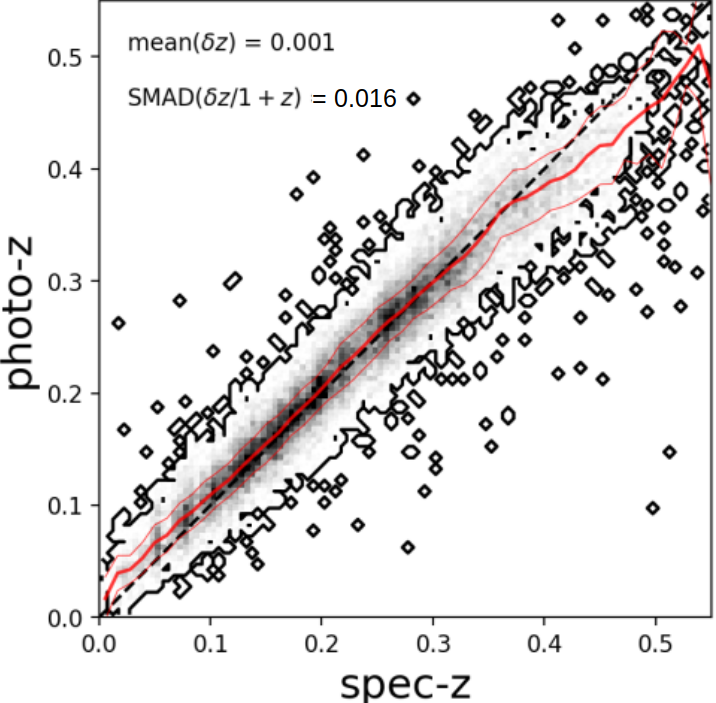}
    \subcaption{Comparison of spectroscopic and the predicted photometric redshifts for the test sample. The thick red solid line represents the running median, while the thinner red lines enclose the scatter, quantified by the SMAD.\label{fig:result}}
        \includegraphics[width=0.9\linewidth]{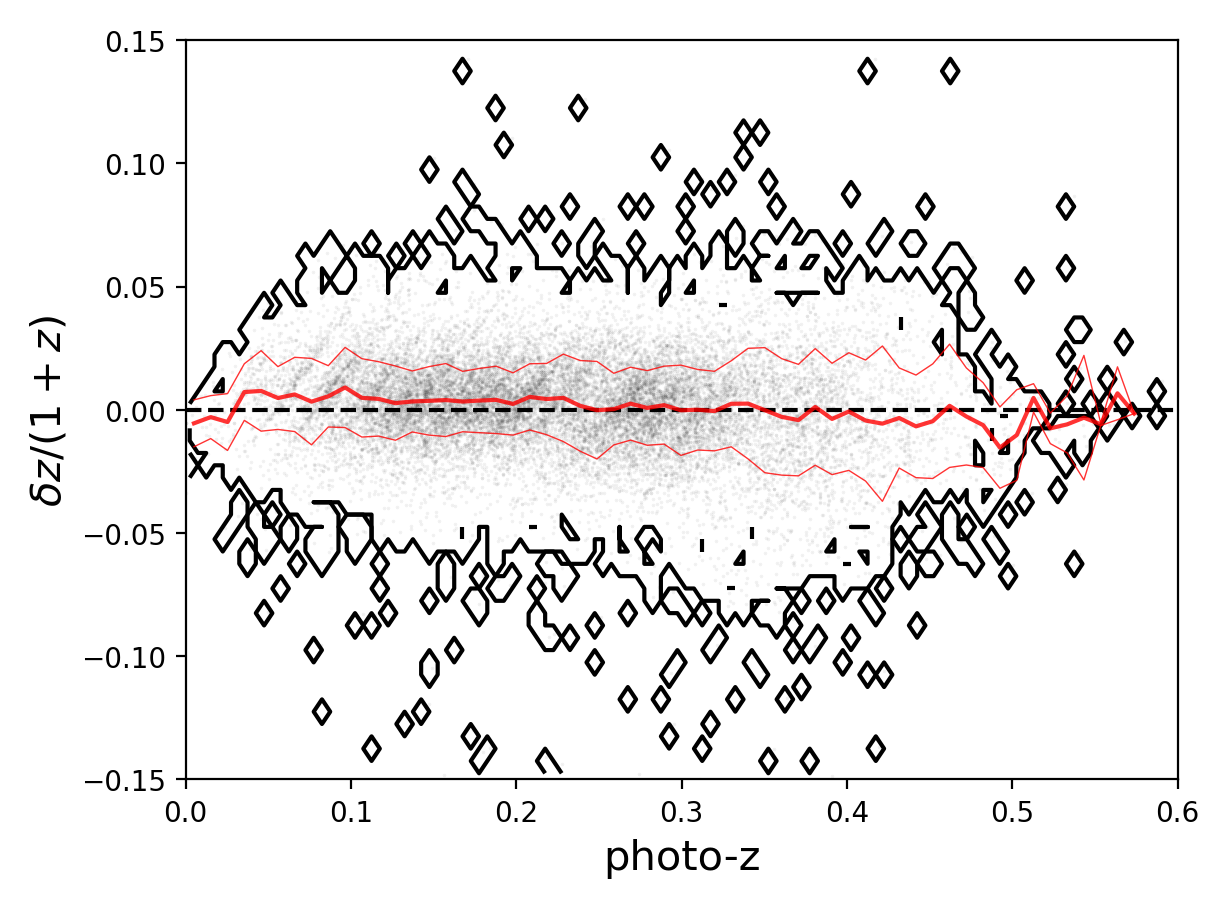}
    \subcaption{Normalised bias as a function of photo-$z$, with red lines encoding the running median and SMAD as in the upper panel. \label{fig:smad}}
  \end{minipage}
\end{figure*}

Our ML model for photo-zs is a combined, multi-input one. Inspired by \cite{Henghes2022,Li2022}, we use both 4-band KiDS images and the corresponding 9-band KiDS+VIKING magnitudes. We combined two different ML schemes: ordinary neural networks (ONN) and CNN (Inception). The architecture of our model is shown in Fig.~\ref{fig:model}.

When tuning the model and assessing its
performance, we primarily employed three metrics: Mean Squared Error (MSE), Mean Absolute Error (MAE), and R-squared error ($R^2$):



\begin{equation}
    R^2 =1 - \frac{\sum_{i=1}^{n} (z_i - \hat{z}_i)^2}{\sum_{i=1}^{n} (z_i - \bar{z})^2} \; ,
\end{equation}
where $n$ is the number of samples used, 
$z_i$ is the predicted value, $\hat{z}_i$ is the true value, and $\bar{z}$ is the mean of true redshift values. $R^2$ is a statistical measure indicating how well a model predicts outcomes in a regression analysis. It represents the proportion of variance explained by the model relative to the total variance. In an ideal scenario, its value is unity, and our combined model has demonstrated the capability to achieve an $R^2$ value exceeding $0.92$. 

MSE is effective for learning outliers, while MAE is advantageous for disregarding them. In our approach, we employed the \cite{Huber1964} loss function that combines the characteristics of both MSE and MAE, providing a balanced approach to handling outliers during the training of the model:

\begin{equation}
    H(z, \hat{z}) = 
    \begin{cases}
        \frac{1}{2}(z - \hat{z})^2, & \text{if } |z - \hat{z}| \leq \delta, \\
        \delta(|z - \hat{z}| - \frac{\delta}{2}), & \text{otherwise}.
    \end{cases}
\end{equation}
The value of $\delta$ we use is 0.0001 by trial and error.

\section{Results}
The photo-z performance is evaluated using the following statistics:
\begin{itemize}
  \item bias: 
  \begin{equation}
       \delta z= z_\mathrm{phot} - z_\mathrm{spec} \, ,
  \end{equation}
   \item normalized bias: 
  \begin{equation}
      \Delta z = \frac{\delta z}{1+z_\mathrm{spec}} \, ,
  \end{equation}
  \item the standard deviation of $\Delta z$,
  \item scaled median absolute deviation (SMAD) of $\Delta z$, where 
\begin{equation}
    \text{SMAD}(x) = 1.4826 \times \text{median} \left(\lvert x - \text{median}(x) \rvert\right) \, .
\end{equation}
 \end{itemize}

These statistics for the overall test sample (based on GAMA equatorial data) are provided in Table \ref{tab:stats}, where they are directly compared with previous results from \cite{Bilicki2021}, derived with the ANNz2 software employing 9-band KiDS+VIKING magnitudes. There is visible improvement in terms of scatter, both in the standard deviation and SMAD, from the earlier ANNz2 results to our new DL model. This shows that our combined model gives promise for improved photo-$z$s for the KiDS Bright Galaxy Sample. We additionally visualize our DL photo-$z$ performance in Figs.~\ref{fig:result} \& \ref{fig:smad}, which show generally unbiased and stable photo-$z$s as a function of the photometric redshift itself.

\begin{table*}[ht]
    \centering
    \caption{\label{tab:stats}Statistics of photometric redshift performance obtained for KiDS-GAMA equatorial spectroscopic sample. }

    \begin{tabular}{cccccc}
    \hline
    \hline\\
       Method & $\langle\delta z\rangle$ & $\langle \Delta z \rangle$ & $\sigma(\Delta z)$ & SMAD$(\Delta z)$ \\
        \hline\\
        ANNz2 \citep{Bilicki2021} & $0.0005$ & $0.0009$ & $0.024$ & $0.018$ \\
        Combined model (this work) & $0.001$ & $0.001$ & $0.021$ & $0.016$ \\
    
      \hline
    \end{tabular}
\end{table*}

\section{Conclusion and Future Prospects}
In these proceedings, we described a new deep-learning model for photometric redshift derivation in the KiDS bright sample that joins Inception-based CNN using 4-band KiDS images, with an ANN employing 9-band magnitudes. We see improvement over previous work by \cite{Bilicki2021}, where ANNs with magnitudes only were used. In the forthcoming paper, we will present more details and apply our model to the entire KiDS DR4 Bright Galaxy Sample, which will give new, improved photo-$z$s for that dataset. 

In the near future, we plan to build a deep-learning model which will use the 9-band imaging from both KiDS and VIKING (unlike KiDS $ugri$ only as presently). That model will be subsequently used to derive photo-$z$s for the Bright Galaxy Sample in the final KiDS Data Release 5 \citep[\textcolor{blue}{Wright et al., in press};][]{Wright2023}.

\section{Acknowledgement}
This work is funded via the research project `Precision and Accuracy for Cosmological Imaging Surveys (PACIS)' from the Polish National Science Centre through a Sonata-Bis grant no.\ 2020/38/E/ST9/00395. We thank Rui Li and Nicola Napolitano for their assistance and feedback.

\bibliographystyle{ptapap}
\bibliography{pta4authors}

\end{document}